\documentclass[conference]{IEEEtran}
\usepackage{cite}
\usepackage[numbers,sort&compress]{natbib}
\usepackage{amsmath,amssymb,amsfonts,gensymb}
\usepackage{algorithmic}
\usepackage{graphicx}
\usepackage{textcomp}
\usepackage{booktabs}
\usepackage{multirow}
\usepackage{xcolor}
\def\BibTeX{{\rm B\kern-.05em{\sc i\kern-.025em b}\kern-.08em
    T\kern-.1667em\lower.7ex\hbox{E}\kern-.125emX}}
\def \am [#1]{\textcolor{blue}{AM: #1}}

\begin{document}

\title{Binaural Signal Representations for Joint Sound Event Detection and Acoustic Scene Classification 
}

\author{\IEEEauthorblockN{Daniel Aleksander Krause, Annamaria Mesaros}
\IEEEauthorblockA{\textit{Faculty of Information Technology and Communication Sciences} \\
\textit{Tampere University}\\
Tampere, Finland \\
daniel.krause@tuni.fi, annamaria.mesaros@tuni.fi}
}

\maketitle

\begin{abstract}
Sound event detection (SED) and Acoustic scene classification (ASC) are two widely researched audio tasks that constitute an important part of research on acoustic scene analysis. Considering shared information between sound events and acoustic scenes, performing both tasks jointly is a natural part of a complex machine listening system. In this paper, we~investigate the usefulness of several spatial audio features in~training a joint deep neural network (DNN) model performing SED and ASC. 
Experiments are performed for two different datasets containing binaural recordings and synchronous sound event and acoustic scene labels to analyse the differences between performing SED and ASC separately or jointly. The presented results show that the use of specific binaural features, mainly the Generalized Cross Correlation with Phase Transform (GCC-phat) and sines and cosines of phase differences, result in a better performing model in both separate and joint tasks as compared with baseline methods based on logmel energies only.

\end{abstract}


\section{Introduction}
Computational auditory scene analysis (CASA) has been a widely researched topic in recent years \cite{Virtanen2018Ananlysis}. Automatic analysis of audio content allows for retrieving information for a wide number of practical applications such as speech recognition \cite{speechrec}, surveillance systems \cite{opatka2011ApplicationOV}, autonomous robots \cite{4058525}, teleconferencing \cite{audiotele} or hearing-impaired support systems \cite{5202720, 6637757}. CASA consists of numerous audio tasks, including sound source localization \cite{krauseconv}, audio tagging \cite{Fonseca2018_DCASE}, sound event detection \cite{Mesaros2019_TASLP} or acoustic scene classification \cite{Mesaros2018a, KeisukeImoto2018E183002}. While most research has been focused on each of the tasks separately, a natural step forward in developing a complex scene analysis system is to create models capable of tackling several interrelated purposes at the same time. This idea has been seen recently in, e.g., research on joint sound event detection and~localization \cite{Adavanne_2019}.

Amongst other tasks, sound event detection and acoustic scene classification share a substantial amount of information. Many sound events are strongly correlated with their scenes, e.g., birds singing are more likely to appear in a forest and~a~park, whereas keyboard typing is most frequently heard in office environments. Over the past years, both tasks have evolved from being based on traditional machine learning techniques like Gaussian mixture models (GMMs) \cite{eventhistograms, 4959522}, hidden Markov models \cite{1561288} or support vector machines \cite{6971128} to utilizing more advanced deep learning techniques \cite{Mesaros2019_TASLP, Mesaros2018a}. The correlation between acoustic scenes and their constituent sound events has been incidentally researched by studying context-aware event detectors in \cite{Heittola2013ContextdependentSE} and \cite{tonami2022sound}. Recently, a~multitask learning (MTL) method for performing both tasks jointly has been proposed by employing a DNN architecture with two separate output branches \cite{tonami2019joint}. The method has been later improved by utilizing dynamic weight adaptation \cite{9689320}. Despite some approaches that included spatial information in~SED \cite{7952260, Adavanne2016} and ASC \cite{Green2017}, most of the current state-of-the-art approaches consist of training DNNs with magnitude-based features like mel-frequency cepstral coefficients (MFCC) \cite{eventhistograms} or plain logmel energies \cite{Martin2021}. 

Since many SED and ASC datasets contain binaural recordings, there is a large unexplored ground for improving both tasks using spatial information. A phasegram is half of~the~signal spectrum, being an important source of information in~many contexts, complementary to the magnitude spectrogram. The addition of phase-derived information gives a~fuller picture about the content of an auditory scene, hypothetically giving room for improvement of a DNN's performance. This difference might become even more pronounced for compound models performing multiple tasks jointly, since they require more information to tackle their higher complexity.

In this paper, we propose several binaural spatial features to improve the performance of a joint model performing sound event detection and acoustic scene classification. We~present a~comparative study, in which  models trained with a feature set including the investigated signal representations are compared with a baseline method based on logmel energies. Experiments are provided separately for SED, ASC and for a joint MTL model and differences in performance between the three options are analysed. The study is performed for two datasets, strongly differing in the size and acoustic content. Finally, we present an analysis of the influence of spatial information on the examined audio tasks altogether with the emerging conclusions. 

The paper is organized as follows. Section \ref{method} describes the~details of the applied method - in \ref{model} we present the~DNN model, and in \ref{features} the analysed audio features. Section \ref{setup} shows the outline of the experiments, including a~description of the utilized datasets in \ref{data-part} and experimental scenarios in \ref{scenario}. The obtained results are presented in~Section \ref{results}, whereas Section \ref{conclusions} sums up the main conclusions of our study.

\section{Method}
\label{method}
\subsection{Model}
\label{model}
\begin{figure}[!t]
    \centering
    \includegraphics[width=80mm,scale=0.8]{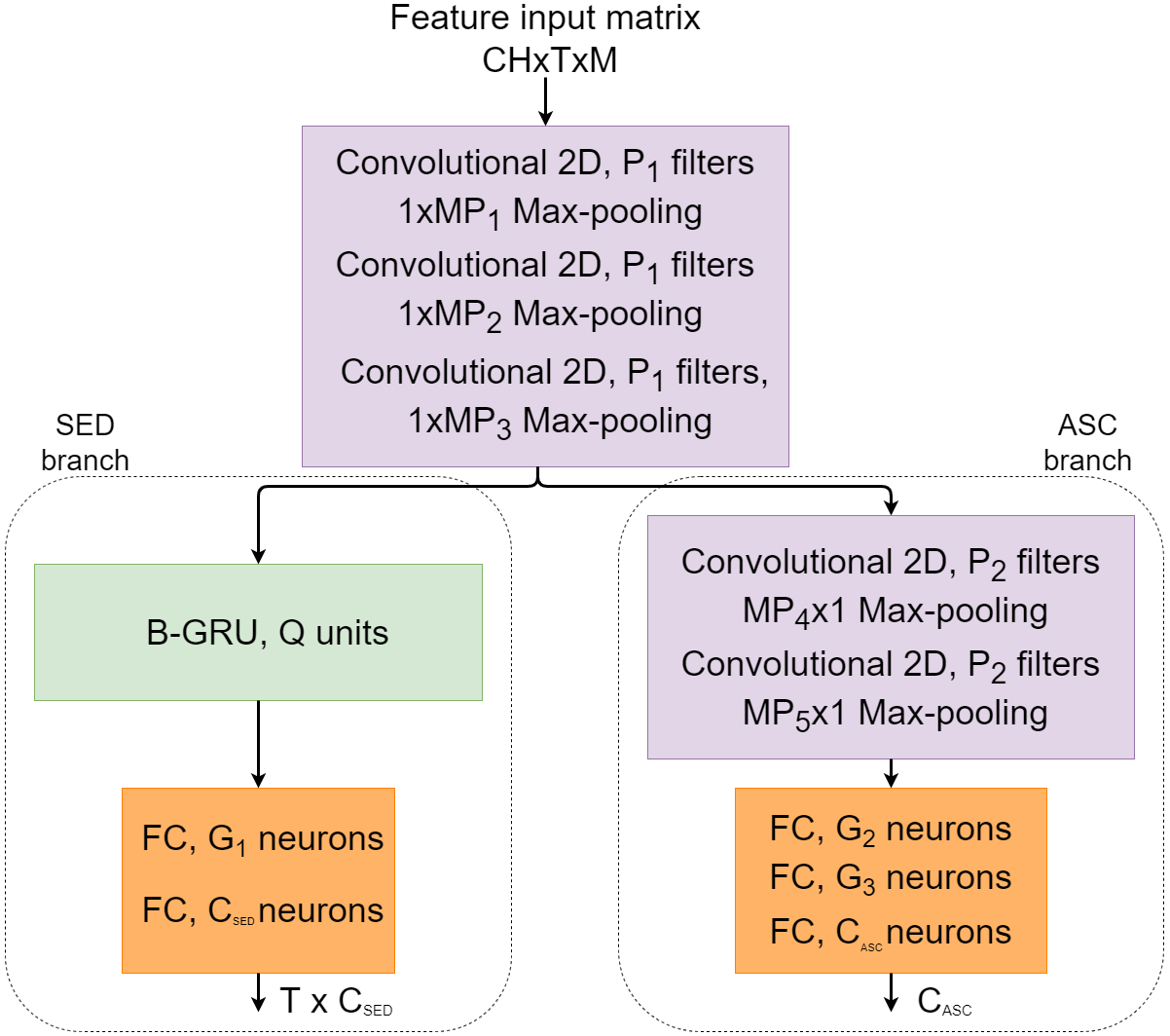}
    \caption{Architecture of the utilized deep neural network.    }
    \label{fig:architecture}
\end{figure}

To enable concurrent performance of event detection and scene classification, we utilize a deep neural network similar to the one proposed in \cite{tonami2019joint}. The DNN architecture is depicted in Fig. \ref{fig:architecture}. A feature input matrix of shape $CH \times T \times M$ is~fed to the model, where $CH$, $T$ and $M$ stand for the number of channels, time sequence length in frames and number of~mel filters respectively. The features are initially processed by~three 2D convolutional layers, each consisting of $P_{1}$ filters and~followed by batch normalization and max-pooling across the feature dimension. The model is then split into two separate branches, one responsible for sound event detection and the other performing acoustic scene classification. 

The detection part consists of a single bidirectional Gated Recurrent Unit (GRU), which allows for temporal modelling of the acoustic signal. $Q$ features are then passed to two fully-connected layers (FC). The first one contains $G_{1}$ linear neurons, where the last layer performs the final modelling of~event onsets and offsets via $C_{SED}$ sigmoidal output neurons, which is equal to the number of sound event classes in the dataset. The second branch of the DNN is further processing the~features with additional two convolutional layers, each with $P_{2}$ filter kernels. Since scene classification scores are returned per clip, each convolutional layer is followed by max-pooling across the $T$ axis to reduce the time dimension. The scene-specific features are fed to three FC layers. The first two are described by $G_{1}$ and $G_{2}$ linear units, whereas the final layer consists of $C_{ASC}$ neurons followed by a soft-max function to~pick only one scene class per clip. In this paper different model parameters are used depending on the utilized dataset, which is further explained in \ref{data-part}. During training, learning weights are applied to both output branches, where the ASC part is weighted with a value 0.0001 as compared with 1~for SED. Models performing SED and ASC as separate tasks are created by simply removing the unnecessary task-specific branch from the joint network. 

\subsection{Features}
\label{features}
To investigate the potential of spatial information for joint sound event detection and acoustic scene classification, we~exploit a number of different features which are suitable for binaural recordings. All experiments are performed by using $M$ \textbf{logmel energies}, which are commonly used in both event detection and scene classification. Firstly, a complex spectrogram is obtained by computing a Short-Time Fourier Transform (STFT) using a Hamming window of 40 ms length and 50$\%$ overlap. Mel energies are calculated based on~the~following formula:\begin{gather}
    \mathrm{\log{|X_{mel}}[n,m]|}=\log{|\sum_{k=0}^{K-1} X[n,k]H_{mel,m}[k]|} ,
\end{gather}
where $X[n,k]$ describes the complex STFT matrix, $H_{mel,m}[k]$ stands for the $m$-th mel filter and $n$ and $k$ are the time and~frequency indices, respectively. To exploit potential variance between binaural channels, we utilize interaural level differences (\textbf{ILDs}) which comprise a major cue in binaural localization \cite{blauert1997spatial} and can be used to add interchannel information above 1.5 kHz. ILDs are calculated using:\begin{gather}\mathrm{ILD}[n,m]=\frac{|X_{mel,l}[n,m]|}{|X_{mel,r}[n,m]|},
\end{gather}
where $l$ and $r$ stand for the left and right channels. 

Apart from magnitude-based features, we investigate several phase-based signal representations. First, raw \textbf{phase} values are used as the most general signal description. To match the~feature dimension of logmel spectrograms, phase values are calculated after applying the same number of mel filters to~the~complex STFT matrix. To express the interaural differences more explicitly, we use the inter-channel phase differences (\textbf{IPDs}) which are defined as:
\begin{gather}
    \mathrm{IPD}[n,m]=\arg(X_{mel,l}[n,m])-\arg(X_{mel,r}[n,m]).
\end{gather}
Next, we compute the sines and cosines of IPDs (\textbf{sin\&cos}):
  \begin{gather}
    \mathrm{SI}[n,m]=\sin{(\mathrm{IPD}[n,m])},
    \\
    \mathrm{CI}[n,m]=\cos{(\mathrm{IPD}[n,m])}.
\end{gather}
Compared with standard phase differences, sines and cosines avoid phase wrapping and show a smoother representation of highly variant phase values. This feature was introduced first in speech separation \cite{sincosfeat} and has been shown to improve performance in acoustic scene analysis \cite{krausefeat}. Finally, we utilize the GCC-phat features (\textbf{GCC}), which are commonly used in~localization-related tasks, including binaural scenarios \cite{brandstein1997robust}. Here, we compute GCC features using the following formula:
\begin{gather}
    \mathrm{GCC}[n,d]=\mathcal{F}^{-1}(\frac{X_l[n,k]\cdot X_r^*[n,k]}{|X_l[n,k]||X_r[n,k]|}),
\end{gather}
where $\mathcal{F}^{-1}\{.\}$ denotes the inverse Fourier transform and~$d$~stands for the time-lag between the channels. In~this~paper, the maximum value of $d$ is cut to $M$ to match the feature dimension of logmel energies as proposed in \cite{cao2019}.

\section{Experimental setup}
\label{setup}
\subsection{Data}
\label{data-part}
We evaluate SED and ASC models on a combination of~datasets, consisting of TUT Sound Events 2016, TUT Sound Events 2017, and Acoustic Scenes 2016 (hereafter referred as \textbf{TUT 2016/2017}), similarly to \cite{tonami2019joint}. This dataset contains 192 minutes of binaural audio files that includes four acoustic scenes: home, residential area, city center and office. In total, the scenes consist of 25 sound event classes: bird singing, brakes squeaking, breathing, car, children, cupboard, cutlery, dishes, drawer, fan, glass jingling, keyboard typing, large vehicle, mouse clicking, mouse wheeling, object banging, object impact, object rustling, object snapping, object squaking, people talking, people walking, washing dishes, water tap running and wind blowing.  

To further investigate some of our results, we provide additional experiments with another dataset - \textbf{TUT SED 2009}, which was first described in \cite{eventhistograms}. 
TUT \nolinebreak SED \nolinebreak 2009 consists of 1133 minutes of binaural audio, recorded in 10 different acoustic scenes: basketball game, beach, inside a~bus, inside a~car, hallway, office, restaurant, shop, street and stadium with track and field events. The scenes contain a total of 61 different sound events (wind, yelling, car, shoe squeaks, etc.) plus one general class, therefore provide a more comprehensive evaluation of the investigated method. Both datasets consist of real life recordings, therefore sound sources can appear on~many different directions.

\subsection{Experiment scenario}
\label{scenario}
We test several feature combinations in order to investigate the influence of spatial information on the performance of~the sound event detection and acoustic scene classifications tasks performed separately and in a joint manner. For all experiments, we treat models trained with a mono logmel spectrogram as a baseline and compare it with the model fed with additional features obtained from both binaural channels. Firstly, we provide a comparison of all features for the TUT 2016/2017 dataset, after which we perform a further investigation of the most promising models using the TUT SED 2009 data. Table \ref{tab:model_parameters} presents the model parameters, chosen depending on the training data.

Models are evaluated using the Error Rate ($ER$) and $F_{1}$~score for the SED part and $F_{1}$ measure for the ASC part. To provide a fair comparison with results available for both datasets in the literature, we use different evaluation methods for different kinds of data. For TUT 2016/2017, we calculate framewise detection scores in 40 ms frames and clipwise scene classification scores, whereas for TUT SED 2009 models are evaluated on a 1-second segment level for SED, and per file for ASC. 
\renewcommand{\arraystretch}{0.7}
\begin{table}[!t]
    \centering
    \caption{Model parameters used for the different datasets.}
    \begin{tabular}{c|cc}
    \toprule
         Parameters & TUT 2016/2017 & TUT SED 2009  \\
         \midrule
         $M$ & 64 & 40 \\
         $T$ & 500 & 1000 \\
         Kernel size & 3x3 & 5x5 \\
         $P$ & [128, 256] & [192, 96] \\
         $MP$ & [8, 2, 2, 25, 20] & [5, 4, 2, 25, 20] \\
         $Q$ & 64 & 128 \\
         $G$ & [128, 512, 256] & [128, 512, 256]\\
         $C_{SED}$ & 25 & 63 \\
         $C_{ASC}$ & 4 & 10 \\
         \bottomrule
         
    \end{tabular}
    \label{tab:model_parameters}
\end{table}
\renewcommand{\arraystretch}{1.0}

\section{Results}
\label{results}

\renewcommand{\arraystretch}{0.7}
\begin{table*}[ht]
\begin{minipage}{\textwidth}
  \centering
  \caption{Detection and classification scores obtained on TUT16/17 data.}
    \begin{tabular}{l|cc|c||ccc}
        \toprule
    \multirow{2}{*}{Features} & \multicolumn{2}{c|}{SED} & \multicolumn{1}{c||}{ASC} &  \multicolumn{3}{c}{MTL} \\
          & $ER$ & $F_{1}[\%]$ & $F_{1}[\%]$ & $ER_{SED}$ & $F_{1,SED}[\%]$ & $F_{1,ASC}[\%]$   \\
    \midrule
    Mel (1ch) & 0.84 ± 0.07  & 35.6 ± 3.2 & 71.9 ± 3.8 & 0.81 ± 0.07 & 36.1 ± 3.2 & 61.0 ± 4.0 \\
    Mel (2ch) & 0.75 ± 0.08 & 43.2 ± 2.9 & 71.1 ± 3.6 & 0.76 ± 0.06 & 41.2 ± 2.5 & 61.7 ± 3.9 \\
    Mel + Phase & 0.81 ± 0.09 & 37.1 ± 3.1 & 53.3 ± 6.1 & 0.76 ± 0.05 & 42.0 ± 2.5 & 66.8 ± 3.5 \\
    Mel + IPD  & 0.75 ± 0.06 & 37.6 ± 3.1 & 63.1 ± 6.5 & 0.79 ± 0.06 & 39.7 ± 3.0 & 60.9 ± 4.1 \\
    Mel + sin\&cos  & \textbf{0.73} ± 0.04 & \textbf{48.4} ± 2.7 & \textbf{73.7} ± 3.6 & \textbf{0.71} ± 0.04 & \textbf{49.9} ± 2.7 & \textbf{68.9} ± 3.0 \\
    Mel + GCC  & 0.77 ± 0.05 & 39.2 ± 3.0 & 70.6 ± 4.0  & \textbf{0.73} ± 0.04 & \textbf{45.4} ± 2.8 & \textbf{70.4} ± 3.2\\
    Mel + ILD  & 0.81 ± 0.06 & 33.2 ± 3.3 & 68.1 ± 4.1  & 0.82 ± 0.06 & 34.0 ± 3.2 & 68.3 ± 3.3\\
    \midrule
    Mel (1ch) \cite{tonami2019joint} & 1.30 & 44.9 & 67.4 & 0.91 & 49.0 & 60.0 \\
    \bottomrule
    \end{tabular}%
  \label{tab:201617results}%
\end{minipage}
  \vspace{-8pt}
\end{table*}%

\begin{table*}[ht]
\begin{minipage}{\textwidth}
  \centering
  \caption{Detection and classification scores obtained on TUT SED 2009 data.
  }
    \begin{tabular}{l|cc|c||ccc}
    \toprule
    \multirow{2}{*}{Features} & \multicolumn{2}{c|}{SED} & \multicolumn{1}{c||}{ASC} &  \multicolumn{3}{c}{MTL} \\
          & $ER$ & $F_{1}[\%]$ & $F_{1}[\%]$ & $ER_{SED}$ & $F_{1,SED}[\%]$ & $F_{1,ASC}[\%]$   \\
    \midrule
    \midrule
    Mel (1ch) & 0.66 ± 0.04  & 54.5 ± 4.2 & 88.2 ± 2.8 & 0.58 ± 0.05 & 58.2 ± 2.4 & 87.8 ± 1.6 \\
    Mel + sin\&cos  & 0.67 ± 0.05 & 53.9 ± 4.7 & 90.5 ± 2.2 & \textbf{0.53} ± 0.03 & 61.8 ± 2.7 & 91.2 ± 1.8 \\
    Mel + GCC  & \textbf{0.64} ± 0.05 & \textbf{55.8} ± 4.3 & \textbf{94.9} ± 2.0  &\textbf{ 0.53} ± 0.04 & \textbf{62.4} ± 2.6 & \textbf{93.1} ± 2.0\\
    \midrule
    Mel (1ch) \cite{emre2017} & 0.48 & 69.3 &- &-&-&- \\
    MFCC \cite{eventhistograms}  & - &-  & 92.4 &-&-&- \\  
    \bottomrule
    \end{tabular}%
  \label{tab:2009results}%
  \end{minipage}
  \vspace{-8pt}
\end{table*}%
\renewcommand{\arraystretch}{1.0}

Table \ref{tab:201617results} presents the results obtained for all feature combinations on TUT 2016/2017 data. The scores are shown for models performing separate tasks (SED, ASC) and in a joint manner (MTL). The scores are the average of five runs and~also include the standard deviation over these runs. 

As can be observed, using logmel energies from both binaural channels instead of one significantly improves the separate sound event detection system, for which the $F_{1}$ score has increased by over 7 p.p in comparison with the logmels from a mono (L+R)/2 version. This might be especially true for~scenarios where certain sound events appear only on one side of the head and the information gets averaged by transforming the binaural information into a single mono channel. As for the use of spatial features, the addition of phase values, IPDs and ILDs to the feature representation has not improved detection scores much. In fact, for ILDs we notice a drop in~$F_{1}$~score, which might suggest that level differences between channels do not provide useful information about sound events. However, noticeable improvements can be seen when using GCC and sin\&cos features. In particular for the latter one, we observe the highest $F_{1}$ score of $48.4\%$ and lowest ER of $0.73$. We relate the usefulness of these features to their smooth representation of phase differences between channels, which allows for efficient capturing of~spatial information. 

For most features, we do not observe significant differences for a separate ASC model. While binaural logmel energies, GCC and ILDs show similar results to a monaural scenario, adding raw phase values or IPDs significantly decrease the~clip-level $F_{1}$ score. However, the use of sin\&cos features along the logmel spectrogram improved the best previous score by around 2 p.p., which while not being a major advance, shows that properly represented phase differences might be~useful also for the ASC task. Both ASC and SED results are on par with the reference numbers presented in \cite{tonami2019joint}. While $F_{1}$ values for SED are slightly lower (with sin\&cos being an exception), $ER$ seem to be consistently lower than in~the~referenced work. The ASC scores also show improvement for most feature combinations. The differences might be~caused by different implementations, learning environments and initial model parameters.

The joint SED and ASC results using an MTL model show that simultaneous modelling of both tasks generally results in an overall decrease of performance. While the scores obtained for SED are mostly similar, ASC scores are in most cases substantially worse than their counterparts obtained with separate models. This is true especially for the model trained with logmel energies only, for which the ASC $F_{1}$~score has decreased by around 10~p.p. for both a monaural and~binaural representation. Interestingly, a smaller decrease is seen for~models trained using additional spatial features. The best results in the MTL scenario are obtained using GCC and sin\&cos features, for which the DNNs obtain $70.4\%$ and $68.9\%$ $F_{1}$ scores, respectively. For the same models we also observe an improvement in the SED branch, by 6.2~p.p. and 1.5~p.p. as compared with a separate model. A notable improvement from using the GCC features suggests that even though spatial features might not significantly improve performance in single-task models, they can still show valuable information for models performing multiple tasks at a time. This might suggest that phase-based features contain some specific inter-task shared information, that can be considered important when studying more complex acoustic models.

Table \ref{tab:2009results} shows the results obtained for the TUT SED 2009 dataset using the three most representative scenarios in our study, namely monaural logmel energies and addition of GCC or sin\&cos features. We performed similar experiments as~for~the~TUT 2016/2017 data. Contrary to the previous dataset, we see no improvement brought by GCC or sin\&cos for the separate SED model; however, significant improvements can be observed for separate ASC. As compared with an $F_{1}$ score of $88.2\%$ for a DNN trained with monaural logmel energies, we see an increase by 2.3 p.p. and 6.7 p.p. for sin\&cos and GCC features, respectively. The results for~the~separate SED model fall in a range which is below the reference $ER$ and $F_{1}$ values presented in \cite{emre2017}, which might be a result of a different DNN architecture and lack of~data-balancing techniques that were used in the former paper. However, the ASC model shows further improvements over the ASC method based on GMMs presented in \cite{eventhistograms} for~the same data.

When performing both SED and ASC in a joint manner, ASC performance is characterized by similar values as compared with the separate model. However, there are notable improvements in the SED branch. For a DNN trained with a logmel spectrogram, the 1-s segment-based $F_{1}$ score improved from $54.5\%$ for a separate model to $58.2\%$ for~the~MTL model. 
Spatial features further improved detection performance, with~the~addition of sin\&cos and GCC features increasing the segment-based $F_{1}$ score by 3.6 p.p. and 4.2 p.p., respectively. 
This improvement might be explained by~the~nature of this dataset, in which sound events are much more context-dependent than for the TUT 2016/2017 data, due to~a~comparably limited number of recording locations, with the strong correlation between the scenes and constituent events enabling the MTL model to make use of the joint information in an efficient way.
In conjunction with the conclusions obtained for the TUT 2016/2017 dataset, this again shows that spatial features provide information which can be exploited particularly when training models performing SED and ASC in a joint task.

\section{Conclusions}
\label{conclusions}
In this paper, we presented a wide overview of binaural features and studied their influence on the behaviour of a DNN model performing joint sound event detection and acoustic scene classification. The results obtained for two different datasets show the significant potential of spatial information to improve complex scene analysis. The addition of spatial information, in particular GCC and sin\&cos features, show improvements for separate SED and ASC models when compared with standard logmel energies. These improvements become more pronounced when performing both tasks in~a~joint manner, showing that spatial information might be particularly important for more complex audio analysis by providing a~more comprehensive view of an acoustic scene. We note that due to a shortfall of datasets enabling concurrent analysis of sound events and acoustic scenes, the results would benefit from further confirmation in future studies using more data.

\section*{Acknowledgment}
The authors wish to acknowledge CSC – IT Center for Science, Finland, for computational resources.


\bibliographystyle{IEEEtran}
\small
\bibliography{conference_101719}

\end{document}